# Non-Equilibrium Spark Plasma Reactive Doping Enables Highly Adjustable Metal to Insulator Transitions and Improved Mechanical Stability for VO$_2$

*Xuanchi Zhou [a]†, Yuchen Cui [a]†, Yanlong Shang [a], Haifan Li [a], Jiaou Wang [b], Ye Meng [c], Xiaoguang Xu [a], Yong Jiang [a]\*, Nuofu Chen [d]\* and Jikun Chen [a]\**

[a] *School of Materials Science and Engineering, University of Science and Technology Beijing, Beijing, 100083, China*

[b] *Beijing Synchrotron Radiation Facility, Institute of High Energy Physics, Chinese Academy of Sciences, Beijing 100049, China*

[c] *Institute for Advanced Materials and Technology, University of Science and Technology Beijing, Beijing 100083, China*

[d] *School of Renewable Energy, North China Electric Power University, Beijing 102206, China*

\*Authors to whom correspondence should be addressed: *jikunchen@ustb.edu.cn* (J. Chen), *yjiang@ustb.edu.cn* (Y. Jiang), *nfchen@ncepu.edu.cn* (N. Chen).

*† X. Zhou and Y. Cui contributed equally to this work.*

**Abstract**

Although vanadium dioxide ($VO_2$) exhibits the most abrupt metal to insulator transition (MIT) properties near room-temperature, the present regulation of their MIT functionalities is insufficient owing to the high complexity and susception associated with $V^{4+}$. Herein, we demonstrate a spark plasma assisted reactive sintering (SPARS) approach to simultaneously achieve *in situ* doping and sintering of $VO_2$ within largely short period (~10 minutes). This enables high convenience and flexibility in regulating the electronic structure of $VO_2$ via dopant elements covering Ti, W, Nb, Mo, Cr and Fe, leading to a wide adjustment in their metal to insulator transition temperature ($T_{MIT}$) and basic resistivity ($\rho$). Furthermore, the mechanical strengths of the doped-$VO_2$ were meanwhile largely improved via the compositing effect of high melting-point dopant oxide. The high adjustability in MIT properties and improved mechanical properties further paves the way towards practical applications of $VO_2$ in power electronics, thermochromism and infrared camouflage.

**TOC Graphic**

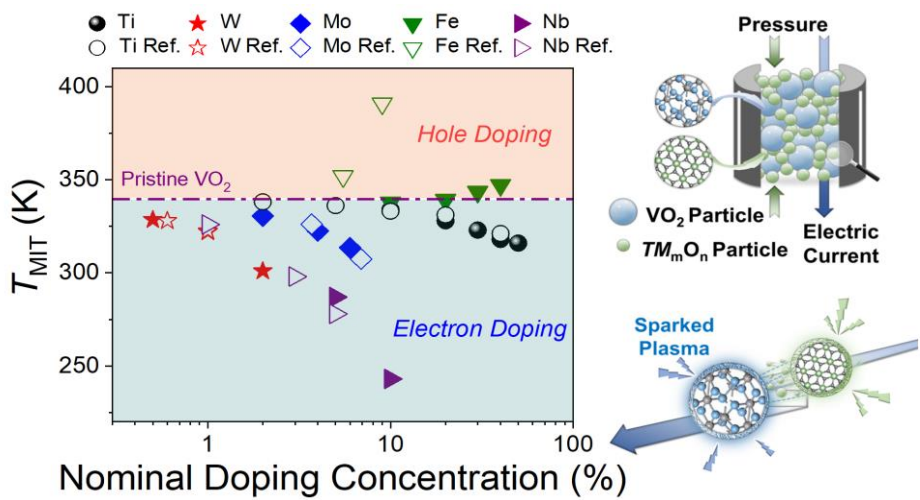

# 1. Introduction

The metal to insulator transition (MIT) within *d*-band correlated oxides brings in abruptions in regulating their electronic and optical properties beyond conventional semiconductors [1-3]. Among the existing family of MIT materials, the vanadium dioxide ($VO_2$) exhibits the most abrupt MIT properties near room-temperature that plays irreplaceable role in applications, such as thermochromism [4-6], infrared camouflage [7,8], and correlated electronic devices [9-12]. Nevertheless, these applications of $VO_2$ are still challenged by its material synthesis as the electronic phase diagram of vanadium oxides is extraordinary complex [13]. For example, the metal to insulator transition temperature ($T_{MIT}$) that triggers the MIT is known to be rather sensitive to the valence state of vanadium that is at the intermediate valence state (e.g., $V^{4+}$) and susceptible to both oxidizing and reducing environment [14-16]. Conventionally, the MIT properties and electronic structure of $VO_2$ is adjusted by the elementary substitution, e.g., the $T_{MIT}$ is elevated or descended via substituting $V^{4+}$ by dopants with lower or higher valence states, respectively [17,18].

To achieve the abrupt variation in the material resistivity during the MIT of $VO_2$, it is vital to precisely control the concentration and homogeneity in the distribution of the dopant elements, without extrinsically altering the oxygen composition [19]. As compared to the vacuum deposition of doped-$VO_2$ thin films via plasmanized or vaporized precursors [20-23], achieving a precise elementary substitution for the bulk materials of $VO_2$ while maintaining their MIT abruption is more difficult. Previously, doped $VO_2$ bulk pellets were made using the following two routes: 1) Firstly synthesizing the powder materials for doped $VO_2$, and afterwards sintering such powder into bulk pellets at precisely controlled atmosphere [24,25]. 2) Co-sintering $VO_2$ and dopant oxides at high temperatures over a long period (e.g., several days) at controlled atmosphere [21]. Usually, the homogeneity and achievable concentration of the dopant element within $VO_2$ are restricted by its equilibrium phase diagram, while the kinetic process that triggers the diffusion of the dopant elements may easily disturb the desired valence state of vanadium [26,27]. Up to date, an effective strategy to accurately adjust the MIT properties of the $VO_2$ bulk material via conveniently achievable elementary substitution is yet lacking, and this impedes applying $VO_2$ for power electronics as discrete elementary devices.

In this article, we demonstrate a spark plasma assisted reactive sintering (SPARS) strategy that largely elevates the flexibility and effectiveness in the elementary substitution of $VO_2$ as bulk material, covering a large abundance of the dopant elements such as Ti, W, Nb, Mo, Cr, and Fe. The non-equilibrium process related to SPARS largely shorten the diffusion period of the dopants (e.g., 10 minutes), and it enables a broad adjustment of the electronic structure and the MIT functionality of $VO_2$. Compliant variations in the electronic structure of $VO_2$ were probed via the near-edge X-ray absorption fine structure (NEXAFS) analysis that differs to the conventional doping of semiconductors. Furthermore, the unreacted dopant oxides with high melting point can be meanwhile used to improve the mechanical properties of $VO_2$, paving the way towards their practical applications in power electronics.

## 2. Experimental Section

A spark plasma assisted reactive sintering (SPARS) strategy was used for synthesizing the doped $VO_2$ bulk materials. The as-proposed SPARS approach includes the following two steps. In the first step, the powder of $VO_2$ is homogenously mixed with the transition metal oxides ($TM_mO_n$) dopants at the desired stoichiometry, and afterwards pressed inside a graphite die. In the second step, a large electric current (e.g., 300 A/cm$^2$) and high pressure (e.g., 20 MPa) are imparted into the graphite die and trigger the solid-state reaction between $VO_2$ and $TM_mO_n$ powders at the temperature of 900 °C for 10 minutes under a vacuum atmosphere. The as-made $VO_2$ bulk material was a pellet with the radius of 10 mm and the thickness of around 4 mm.

The crystal structure of samples was characterized by the X-ray diffraction (XRD) (RIGAKU, Smartlab). The valence state of as-made samples was characterized by X-ray photoelectron spectroscopy (XPS) (Kratos, AXIS ULTRA$^{DLD}$) that was equipped with monochromatic Al-Kα source (150 W) and photon energy of 1486.6 eV. All the binding energies as obtained from the XPS were referenced to the C 1s peak from the adventitious carbon at 284.8 eV, while a Shirley function was used to subtract the background. In addition, the valence state of vanadium for $VO_2$ is indicated by the binding energy of V $2p_{3/2}$ and V $2p_{1/2}$ peaks. To fit the XPS curves, the V $2p_{3/2}$ peak can be resolved by the two binding energies at 516.3 and 515.7 eV that represents the valence state of $V^{4+}$ and $V^{3+}$, respectively. The V $2p_{1/2}$ region can be analogously divided into the two peaks that are respectively located at 523.6 and 523.0 eV, according to the previous reports [28, 29]. And their morphologies were characterized by scanning electron microscopy (SEM) and energy dispersion spectrum (EDS). The electronic structures of as-made samples were also investigated by the near edge X-ray absorption fine structure (NEXAFS). To characterize their electrical transportation properties, their temperature dependences of resistivity within 300-400 K were measured by using a commercialized CTA-system in vacuum, while the one within 1.8-400 K was measured by using the physical property measurement system (PPMS) (Quantum Design). The Vickers hardness of $VO_2$ bulk material were measured by the digital Vickers hardness tester (VTD 512) at room-temperature under the load of 500 kgf.

## 3. Results and Discussion
### 3.1. Strategy for elementary substitution of $VO_2$ via SPARS

To address the center challenge in the precise element doping for bulk $VO_2$, herein a spark plasma assisted *in situ* doping strategy is used, as illustrated in Fig. 1a. Unlike the previous literatures that uses the reactive spark plasma process for sintering the powder material into bulk pellet, or directly synthesizing compounds from their respective elementary substances [30, 31], herein the reactive spark plasma is used to achieve the element doping of oxides. Firstly, the powder of $VO_2$ is homogenously mixed with the doped transitional metal oxides ($TM_mO_n$, $TM$ = Ti, W, Nb, Mo, Fe, and Cr) and pressed inside a graphite die. Afterwards, a large DC electric current (e.g., around 300 A/cm$^2$) and high pressure (e.g., 20 MPa) are imparted to spark over the powder precursor that triggers their mutual element diffusion and rapidly elevate the temperature to 900 °C (e.g., 60% of the melting point of $VO_2$) for their reactive sintering into bulk pellet. Such spark plasma assisted non-equilibrium process brings in the electrical break down process within the sintering that promotes the elementary diffusion and shortens the period associated with the reactive sintering (e.g., 10 minutes). Meanwhile, the residual of the dopant oxides with high melting point within the grain boundary can be used to improve the mechanical properties of the $VO_2$ pellet. Furthermore, as-proposed SPARS approach makes it possible to flexibly adjust the dopant composition and concentration within $VO_2$ with high convenience, by just using the powder of pure $VO_2$ and the dopant oxides as precursors.

### 3.2. Metal to insulator transition properties for Ti substituted $VO_2$

As-proposed SPARS approach was first applied to achieve Ti-doping of $VO_2$, and Fig. 1b shows the X-ray diffraction (XRD) patterns of as-made $V_{1-x}Ti_xO_2$ pellets. In general, the $V_{1-x}Ti_xO_2$ exhibit the same crystal structure to $VO_2$ (M), while their lattice expands with an increasing $x$ as demonstrated in Fig. 1c (see more details of their structure information from the Rietveld refinement in Table S1). It is also worth noticing that the $c_0/a_0$ in the lattice constant slightly increases indicated by the Rietveld refinements (see Fig. S1), and this observation is in agreement to the understanding by C. N. R. Rao et al that a local structural distortion upon Ti substitution may also relevant to the variation in MIT properties[32]. This clearly demonstrates the effective substitution of the lattice vanadium within $VO_2$ by titanium that exhibits larger ionic radius, as is in agreement to the previous report [33]. The cross-section morphology of as-sintered $V_{1-x}Ti_xO_2$ pellet as shown in Fig. 1d demonstrates its polycrystallinity with a grain size of ~20 μm, and the same distribution is observed for V and Ti as further confirmed by the energy dispersion spectroscopy (EDS) shown in Figs. 1e and 1f. Furthermore, X-ray photoemission spectroscopy (XPS) analysis shown in Fig. 1g indicates the generation of $V^{3+}$ valence state via Ti doping of $VO_2$, while the valence state of the Ti remains as 4+ as shown in Fig. 1h. The above understanding about the presence of $V^{3+}$ valence state via Ti doping is further confirmed by the XPS results for other Ti-substituted samples, as shown in Figs. S2 and S3. Fig. 1i demonstrates the temperature dependence of resistivity ($\rho$-$T$) as measured for as-made $V_{1-x}Ti_xO_2$ pellets, where abrupt reductions in their resistivity are observed in $\rho$-$T$ tendencies by elevating temperature across their $T_{MIT}$. The elevation in the resistivity for the Ti-substituted $VO_2$ is against the conventional expectation that the presence of $V^{3+}$ should strengthen the metallic phase that reduces the room-temperature resistivity. It is more likely to be associated with the compositing effect of the residual $TiO_2$ remaining at the grain boundary.

Nevertheless, this is hard to be distinguished in the XRD or EDS mapping, since the particle size of $TiO_2$ (∼ 10-20 nm) is much smaller than $VO_2$ (∼ 20-30 μm) while the crystal structure is also similar for $TiO_2$ and $VO_2$. Similar elevation in the room-temperature resistivity was also previously observed in ref 32, and was attributed to local structure distortion. Therefore, the mechanism related to the Ti doping in $VO_2$ is expected to be more complicated that is worthy to be further explored in the future. In Fig. 1j, the $T_{MIT}$ of as-made $V_{1-x}Ti_xO_2$ pellets were further plotted as a function of doping concentration. As shown $T_{MIT}$ was the averaged ones from the $R$-$T$ tendency measured by heating up and cooling down, while each critical temperatures were taken when the temperature coefficient of resistance ($TCR$) reached the maximum, as illustrated in Fig. S4 and Table S2. It can be seen that as-achieved $T_{MIT}$ exhibits a generally reducing tendency with an increasing substitution concentration of Ti. This observation is in general agreement to the previous report [32], and this is attributed to the distorted local structure[32] and the presence of $V^{3+}$ valence state induced by the substitution of titanium. We emphasize that the Ti substitution of $VO_2$ as achieved presently via our SPARS approach reduces the $T_{MIT}$ that is relevant to the Ti doping composition, while the underneath mechanisms could be complicated as is indeed under debate yet according to the previous reports.

### 3.3. Extending the SPARS strategy to achieved broad regulation in $T_{MIT}$ of $VO_2$

To further extend the regulation of $T_{MIT}$, more dopant oxides (e.g., $WO_3$, $MoO_3$, $Nb_2O_5$, $Cr_2O_3$ and $Fe_2O_3$) were co-sintered with $VO_2$ via the as-proposed SPARS approach, and their respective XRD pattern and representative morphology are shown in Fig. S5 and S6, respectively. It can be seen that the as-synthesized polycrystalline ceramic pellets via the spark plasma assisted sintering process exhibit relatively densified morphology without obvious porosity, and this is further confirmed by its pronounced transition abruption across the $T_{MIT}$. It can be seen that the as-synthesized polycrystalline doped $VO_2$ ceramic pellets via the spark plasma assisted sintering process exhibit relatively densified morphology, despite the presence of little amount of pores, as indicated in Fig. S7. The as-observed morphology is in consistency with the previous reports on polycrystalline $VO_2$ bulk pellets without doping as synthesized by solid-state reaction or spark plasma sintering [34, 35]. This indicates that the as-proposed SPARS strategy can effectively achieve the reactive sintering process that sinters the precursor powders into densified bulk pellet within largely shortened reaction period of 10 minutes. As their $\rho$-$T$ tendencies shown in Fig. 2a, the $T_{MIT}$ is more broadly adjusted within 240-350 K. In Fig. 2b, as-achieved $T_{MIT}$ were plotted as a function of the substitution concentration, and further compared with the previous reports [21, 32, 36-38]. Similar to the previous reports for doped $VO_2$ samples [19], herein substituting the vanadium with elements at higher valence state (e.g., $W^{6+}$, $Mo^{6+}$ and $Nb^{5+}$) reduces the $T_{MIT}$, while the respective substitution via lower valence elements (e.g., $Cr^{3+}$ and $Fe^{3+}$) slightly elevates the $T_{MIT}$. It is worth noticing that apart from the variation in the valence state of vanadium, the variation in the lattice constant owing to the different ionic radius of the dopants compared to vanadium may also influence the MIT properties. Nevertheless, in the present work, all the dopant elements (e.g., $Ti^{4+}$, $Nb^{5+}$, $W^{6+}$, $Mo^{6+}$, $Cr^{3+}$ and $Fe^{3+}$) enlarge the lattice constants of $VO_2$ as further summarized in the Table S3, but $Cr^{3+}$ and $Fe^{3+}$ increase the $T_{MIT}$ of $VO_2$, while $Nb^{5+}$, $W^{6+}$ and $Mo^{6+}$ reduce its $T_{MIT}$. This indicates that the regulation in the valence state of vanadium should be a dominant cause to $T_{MIT}$. In Fig. 2c, the abrupt variations in the resistivity across the MIT ($\rho_{Insul.}$ / $\rho_{Metal.}$) is further plotted as a function of $T_{MIT}$, where the magnitude generally decays with their $T_{MIT}$. It is in particular worth noticing that high transition abruption is achieved for Ti substituted $VO_2$ (e.g., $\rho_{Insul.}$ / $\rho_{Metal.}$ exceeding $10^2$) that is comparable to the presently commercial critical temperature resister (CTR) as reported previously [39].

To evaluate the ability in regulating the $T_{MIT}$ for various dopant elements, the variation in $T_{MIT}$ is linear fitted with the doping concentration (see Fig. S8), and in Fig. 2d as-obtained slope ($dT_{MIT}$ / $dn_{dopant}$) is further plotted as a function of the dopant valence state. It can be seen that the variation in $T_{MIT}$ is more effectively reduced by substituting vanadium with the dopant elements at higher valence states (e.g., -17.9 K %$^{-1}$ for $W^{6+}$ and -9.4 K %$^{-1}$ for $Nb^{5+}$). In contrast, the regulation in $T_{MIT}$ is less effective when the substituting elements exhibits the same or lower valence state compared to $V^{4+}$ (e.g., +2.6 K %$^{-1}$ for $Fe^{3+}$ and -0.43 K %$^{-1}$ for $Ti^{4+}$).

### 3.4. Compliant variation in electronic structures as probed by NEXAFS

To further qualitatively probe the variations in the electronic structures of $VO_2$ associated with the elementary substitution, the near-edge X-ray absorption fine structure (NEXAFS) analysis was performed, as the results shown in Fig. 3a-3b for V-*L* edge and O-*K* edge, respectively. In the V-*L* spectrum of $VO_2$, the V-$L_{III}$ peak (~518 eV) is recognized to be associated with the V $2p_{3/2} \rightarrow 3d$ transition, while the V-$L_{II}$ peak (~524 eV) reflects the V $2p_{1/2} \rightarrow 3d$ transition [40-42]. From the perspective of the hybridization between V 3*d* and O 2*p* band, the peaks in the O-*K* edge (e.g., O 1*s* $\rightarrow$ 2*p* transition) were respectively corresponding to the $t_{2g}$ and $e_g$ orbitals, and the ratio in intensities of $t_{2g}$ / $e_g$ was reported to indicate the orbital occupancy [43, 44]. Furthermore, the $d_{//}$ orbital associated with the neighboring V ions along the rutile c axis reflects the V-V interaction, while the π* orbit pointed in the ligands is corresponding to the electron density that is shifted away from the V-O bond direction. In contrast, σ / σ* orbits with rotational symmetry around V-O bond is associated to the V-O interaction [45].

Substituting the vanadium with elements at higher valence state (e.g., $W^{6+}$, $Mo^{6+}$ and $Nb^{5+}$) results in the leftwards shift of the V-*L* edge (see Fig. 3a), and this observation is in consistency with the XPS results shown in Fig. S9 (e.g., Fig. S9 a and S9 c) that demonstrate the reduction in the valence state of vanadium towards $V^{3+}$.

In addition, the as-observed similar peak position of O 1s peak (e.g., ~530.1 eV) for doped $VO_2$ with respect to the pristine one indicates the main presence of lattice oxygen within the material. Meanwhile, the ratio in intensity of $t_{2g}/e_g$ is observed to be smaller compared to the pristine $VO_2$, and this demonstrates an enhanced orbital occupancy in $t_{2g}$ band, owing to the electron doping from the higher valence substitution. It is known for $VO_2$ that electron carrier doping strengthens the metallic phase and reduces $T_{MIT}$, as illustrated in Fig. 3c [20]. Therefore, the as-observed reduction in the $T_{MIT}$ via high-valence substitution is associated to the electron occupancy in the $t_{2g}$ band of $VO_2$. In addition, the as-observed reduced intensity of the $t_{2g}$ peak with respect to the $e_g$ peak also indicates the weakening of direct V-V interactions upon the cationic substitution [45]. It is also interesting to note that similar variations in the electronic structure are observed for Ti substituted $VO_2$, in which case the V-$L$ edge also shifts leftwards while the $t_{2g}/e_g$ ratio decreases. The above observation is in agreement to the reduction in $T_{MIT}$ for Ti substituted $VO_2$, although the dopant ($Ti^{4+}$) exhibits the same valence state compared to the matrix ($V^{4+}$). It is worth noticing that the reduction as observed in the $T_{MIT}$ of Ti doped $VO_2$ is consistency with the previous reports [32, 46]. Nevertheless, the underneath mechanism associated to the regulation in the $T_{MIT}$ via equivalent substitution is still in debate. For example, C. N. R. Rao *et al* ascribed it to the distorted crystal structure induced by the Ti substitution [32], while I. K. Kristensen attributed such result to the variation in carrier concentrations [46].

In contrast, substituting vanadium with dopant at lower valence state (e.g., $Fe^{3+}$) results in opposite variations in the vanadium valence state, as the V-$L$ edge shifts rightwards in the V-$L$ edge spectrum. This is in agreement to the XPS result that the valence state of vanadium was slightly enhanced towards higher valence (e.g., +5) valence (see Fig. S9 e), and this is also in general consistency to the observed elevation in $T_{MIT}$, as illustrated in Fig. 3c. Nevertheless, it is worth noticing that the relative intensity of $t_{2g}$ peak, compared to $e_g$, in the O-$K$ edge is not increased as expected for lower valence element substitution of vanadium, but also reduced. This reveals the presence of a second mechanism that may also influence the electronic structure of the presently made doped $VO_2$ samples, in addition to the substitution of vanadium. This is most likely to be associated to the generation of oxygen vacancy during the SPARS that will also contribute electron carrier from the perspective of anion regulation [47], and as a result induce the elevation in $T_{MIT}$ via Fe substitution, as compared to the previous reports [38]. In addition, the oxygen vacancy within $VO_2$ is reported to be formed upon a high-temperature vacuum atmosphere, and this is also the case during the SPARS process in coarse vacuum atmosphere [48]. It is also worthy to mention that such potentially generated oxygen vacancy is not the dominate reason for the Ti substitution or higher valence element substitution of vanadium, since a monotonic variation in $T_{MIT}$ with the substitution concentration is indeed observed for all these samples sintered at the same condition. Furthermore, the expected $T_{MIT}$ (around 340 K) and variation in resistivity across MIT (over $10^2$) similar to the previous reports were achieved in as-sintered $VO_2$ pellet without any doping, and this profoundly demonstrate that the potential generation in oxygen vacancy is not problematic in the present work. From the NEXAFS analysis, it is noteworthy that in contrast to conventional semiconductive doping process that mainly alters the carrier concentration, the elementary substitution of $VO_2$ also results in compliant variation in their electronic structures.

**3.5. Meanwhile improving the mechanical strength of the substituted $VO_2$**

Apart from regulating the electronic structures and MIT properties, it is also worth noticing that the mechanical properties of $VO_2$ pellet can be meanwhile improved along with the elementary substitution process using the SPARS approach via compositing with the residual dopants. Notably, the abrupt variation in the lattice constant of $VO_2$ across the $T_{MIT}$ is the main problem that interrupts their electronic functionality in potential applications as critical temperature resister (CTR). For example, the structural variation promotes the generation and propagation of the cracks within $VO_2$ that reduce the abruption in their MIT and deteriorate the mechanical strength. Therefore, both the MIT properties and the mechanical strength were expected to be improved for the practical electronic applications of $VO_2$.

Fig. 4a compares the Vickers hardness ($H_V$) measured for doped $VO_2$ pellets herein made by SPARS method as a function of the doping concentration. It can be seen that the $H_V$ for doped $VO_2$ (e.g., over 500 kgf / $mm^2$) largely exceeds the one observed for the pristine $VO_2$ (e.g., 267.2 kgf / $mm^2$), while $H_V$ shows a generally increasing tendency with the doping concentration. This is attributed to the presence of the residual dopant oxide was not fully diffuse into the lattice of $VO_2$ during the SPARS process that strengthens the grain boundary. In Fig. 4b, we further summarize the achievable mechanical strength for the doped $VO_2$ as made in this work plotted as a function of their transition abruption in resistivity across the $T_{MIT}$. It clearly demonstrates the feasible type and composition in the dopant oxides to achieve abrupt MIT functionality while maintaining good mechanical strength (marked by the square in Fig. 4b). As-observed mechanical strength in the $VO_2$ pellet composited with the high melting-point oxides were comparable to the one observed for $VO_2$ microcrystal as measured by the nanoindentation testing [49]. It is worth noticing that the grain boundary within polycrystalline $VO_2$ bulk material inevitably leads to the suppression in its macro mechanical property, as compared with the nanoindentation associated to the nanoscale hardness.

In particular, the co-sintering of $VO_2$ with dopant oxides exhibiting high melting point (e.g., $Al_2O_3$, $HfO_2$ and $CoO$) not only improve the mechanical strength of the pellets, but also enhance the transition abruption in their $\rho$-$T$ tendencies across $T_{MIT}$. This is demonstrated in Fig. 4c, where more abrupt MIT functionality with similar $T_{MIT}$ is observed for Hf, Al and Co doped $VO_2$ pellet compared to the pristine $VO_2$ pellet. In particular, the magnitude of their basic resistivity at room-temperature ($\rho_{300 K}$) can be regulated within a broad range across two orders of magnitudes as demonstrated in Fig. 4d, and this is expected to extend the thermistor functionality. As a representative, Fig. 4e demonstrates the distribution of the elementary distribution of V, Co and O as measured by energy dispersion spectroscopy (EDS) for co-sintered $VO_2$ and CoO with a nominal composition of $V_{0.7}Co_{0.3}O_2$. It can be seen that the cobalt element is not observed in the region associated with the vanadium dioxides, and this differs to the situation for the aforementioned dopant oxides with low melting points (e.g., $WO_3$, $Nb_2O_5$, $TiO_2$, etc.).

The above results indicate the less effective diffusion of Co within $VO_2$ during the sintering process owing to the relatively high melting point of CoO (see more details in Fig. S10), and this is in agreement to the small variation in $T_{MIT}$ as observed in Fig. S11. This understanding is further confirmed by the Rietveld refinement of their XRD patterns as shown in Fig. S12, while their corresponding percentage of residual dopants (e.g., $Al_2O_3$, CoO, $Nb_2O_5$, $WO_3$ and $TiO_2$) were further summarized in Table S4. It demonstrates that the reaction ration ($P_{react}$) between the dopant oxides with low melting point ($T_{melt}$) and $VO_2$ (e.g., $P_{react}$ is of almost ~ 99.6 % for $Nb_2O_5$ with $T_{melt}$ of 1485 °C) largely exceeds the ones observed for the dopant oxides with high $T_{melt}$ (e.g., $P_{react}$ is of ~ 80 % for CoO with $T_{melt}$ for 1935 °C). Nevertheless, the residual dopant oxides, such as $Al_2O_3$, $HfO_2$ and CoO, significantly improve the mechanical properties of $VO_2$ without sacrificing the MIT properties. From the perspective of mechanical properties, the presence of compositing oxides at high insulating state (e.g., $Al_2O_3$, $HfO_2$ and $Co_3O_4$) can effectively prohibit the generation and propagation of the cracks along the grain boundary. Therefore, a more stable electronic functionality for the as-made doped $VO_2$ bulk material is expected, which is a critical issue for the practical applications in electronic devices upon thermoshock. From the perspective of MIT properties, the compositing with the insulating oxides located at the grain boundary can reduce the charge leakage along the cracks within grain boundary for the insulating phase of $VO_2$, and this leads to abrupt transitions in resistivity across the $T_{MIT}$. Similar effects were also previously reported for the vanadium phosphate glass (VPG) [50] and vanadium-based critical temperature resistor [39]. Therefore, a more stable electronic functionality for the as-made doped $VO_2$ bulk material is expected, which is a critical issue for the practical applications in electronic devices upon thermoshock [51].

## 4. Conclusions

In summary, we largely improved the effectiveness in regulating the MIT functionality via elementary doping as well as the mechanical properties of $VO_2$ bulk material by developing a spark plasma assisted reactive sintering (SPARS) approach. Compared to the conventional solid-state reactions, the spark plasma induced non-equilibrium process promotes the diffusion of the dopant element from its oxide precursor into the lattice of the powder of $VO_2$ within a largely shortened co-sintering period of ~10 minutes. This brings in high flexibility and convenience to control the dopant type and concentration within bulk $VO_2$, enabling a broad adjustment in the $T_{MIT}$ (e.g., within 240-350 K). In contrast to conventional semiconductive doping process, compliant variations in the electronic structure are observed within the doped $VO_2$ as probed by NEXAFS, and therefore variation in the MIT properties is more likely to be associated with the doping-induced electronic orbital regulation rather than simply varying the carrier concentration. Furthermore, the mechanical strength of the doped $VO_2$ bulk material can be meanwhile largely improved (e.g., from 267.2 to 950 kgf / $mm^2$), owing to compositing effect of the unreacted high melting-point oxides. The high tunability in the MIT functionality and improved mechanical strength as enabled by the SPARS approach pave the way towards practical applications of $VO_2$ as a bulk material in power electronics.

**Notes**

The authors declare no competing financial interest.

**Acknowledgements**

This work was supported by the National Key Research and Development Program of China (No. 2021YFA0718900), the National Natural Science Foundation of China (No. 62074014 and 52073090), and the Beijing New-star Plan of Science and Technology (No. Z191100001119071).

**Supporting Information**

Supporting information is available online. See the supporting information for more details about the synthesis method, the derivation of $T_{MIT}$, X-ray diffraction (XRD) patterns, the energy dispersion spectrum (EDS), X-ray photoelectron spectroscopy (XPS), the Rietveld refinements and additional analysis results.

**Correspondences:** Correspondences should be addressed: Prof. Jikun Chen (jikunchen@ustb.edu.cn), Prof. Yong Jiang (yjiang@ustb.edu.cn) and Prof. Nuofu Chen (nfchen@ncepu.edu.cn).

**Figures and captions**

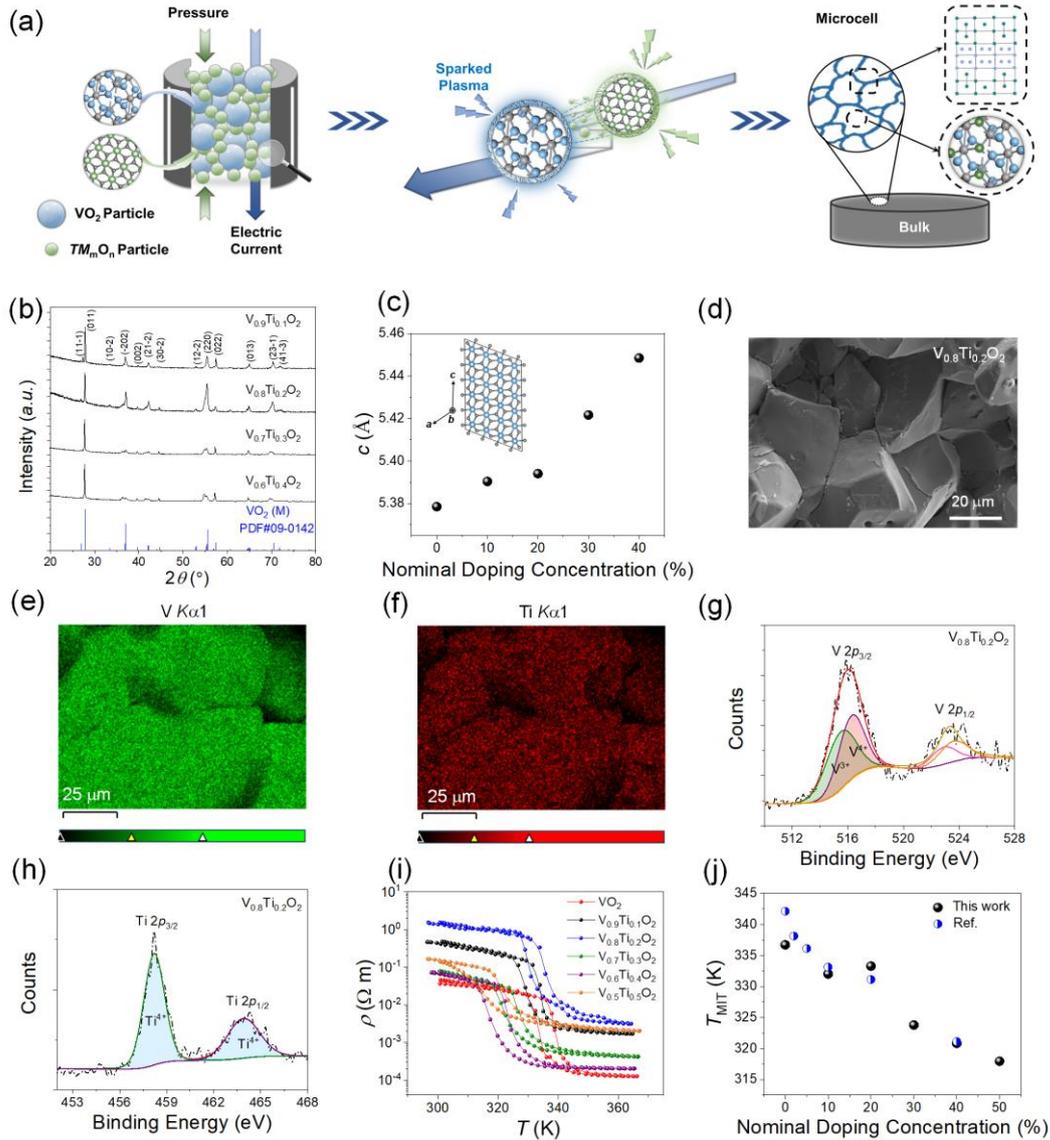

**Fig. 1.** (a) Schematic illustration of as-proposed spark plasma assisted reactive sintering (SPARS) doping strategy that achieves the elementary doping and grain boundary modifications of the bulk $VO_2$. (b) X-ray diffraction (XRD) patterns of as-made $V_{1-x}Ti_xO_2$ bulk material (x from 0.1 to 0.4), as indexed to the structure of the $VO_2$ (M) with the $P2_1/c$ space group. (c) The $c$-axis lattice constants of as-made $V_{1-x}Ti_xO_2$ plotted as a function of x. (d) The cross-section surface morphology of as-synthesized $V_{0.8}Ti_{0.2}O_2$ bulk material from scanning electron microscopy (SEM). (e), (f) The energy dispersion spectrum (EDS) of (e) V and (f) Ti elements for the as-made $V_{0.8}Ti_{0.2}O_2$ bulk. (g), (h) The X-ray photoemission spectroscopy (XPS) analysis of (g) V 2$p$ peak and (h) O 1$s$ peak for the $V_{0.8}Ti_{0.2}O_2$ bulk. (i) Temperature dependence of the resistivity for as-made $V_{1-x}Ti_xO_2$. (j) The metal to insulator transition temperature ($T_{MIT}$) of as-made $V_{1-x}Ti_xO_2$ plotted as a function of the nominal Ti substitution concentration, compared with the previous report [32].

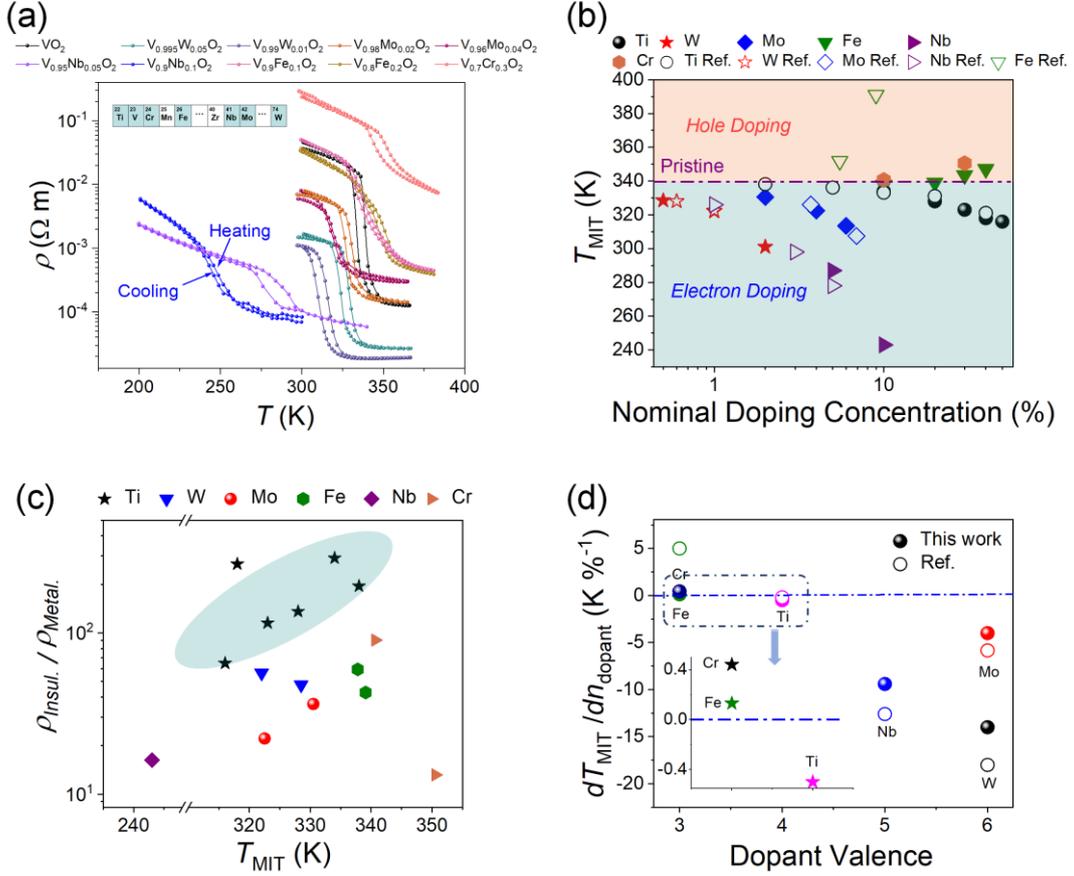

**Fig. 2.** (a) The temperature dependence of the resistivity for as-made transitional metal (*TM*) doped $VO_2$ pellet (*TM* = Fe, W, Nb, Cr and Mo). (b) The $T_{MIT}$ of as-made doped $VO_2$ plotted as a function of the nominal doping concentration and also compared to the previous reports [19, 22, 32, 36, 38]. (c) The variation in the resistivity across the metal to insulator transition ($\rho_{Insul.}/\rho_{Metal.}$) of the doped $VO_2$ made in this work, plotted as a function of their metal to insulator transition temperature ($T_{MIT}$). (d) The modulation capability in $T_{MIT}$, as evaluated by $dT_{MIT}/dn_{dopant}$, for various *TM* dopant elements plotted as a function of their valence state, as also compared with the previous reports [19, 22, 32, 36, 38]. The inset shows the zoom in results for Ti, Cr and Fe.

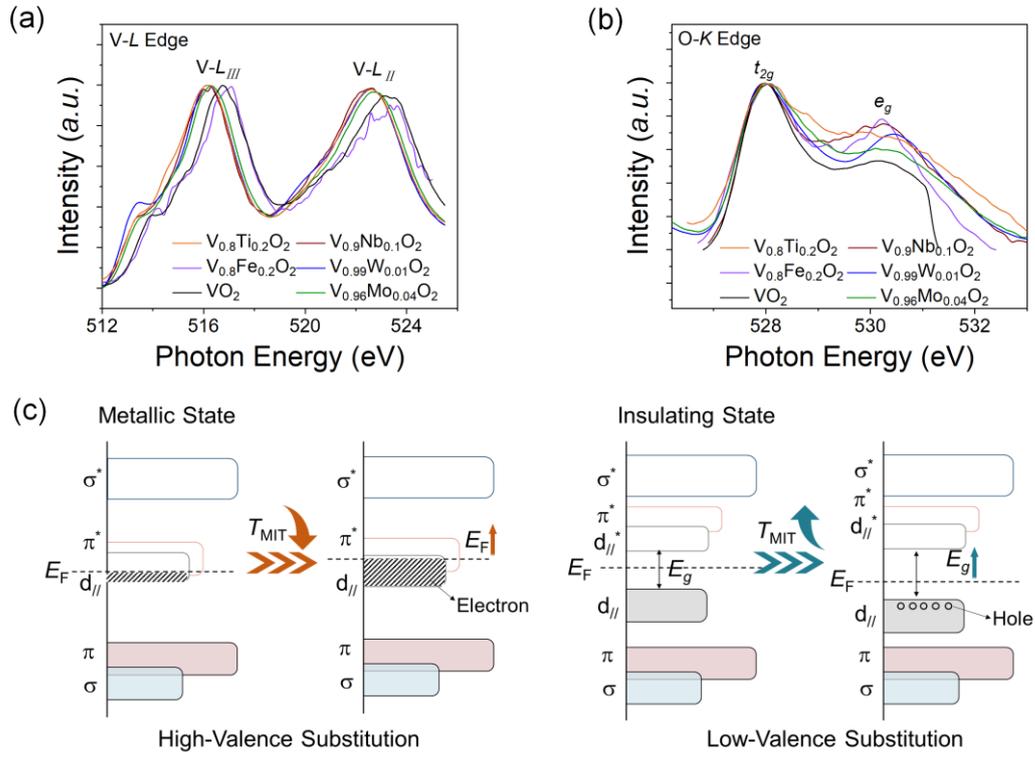

**Fig. 3.** (a), (b) The near-edge X-ray absorption fine structure (NEXAFS) analysis of (a) V-*L* edge and (b) O-*K* edge for the transitional metal (*TM*) substituted $VO_2$ (*TM* = Ti, Fe, W, Nb and Mo), compared to the pristine $VO_2$. (c) Illustrating the variation in electronic structures of $VO_2$ via high-valence (e.g., W, Nb and Mo) and low-valence (e.g., Fe and Cr) *TM* substitutions.

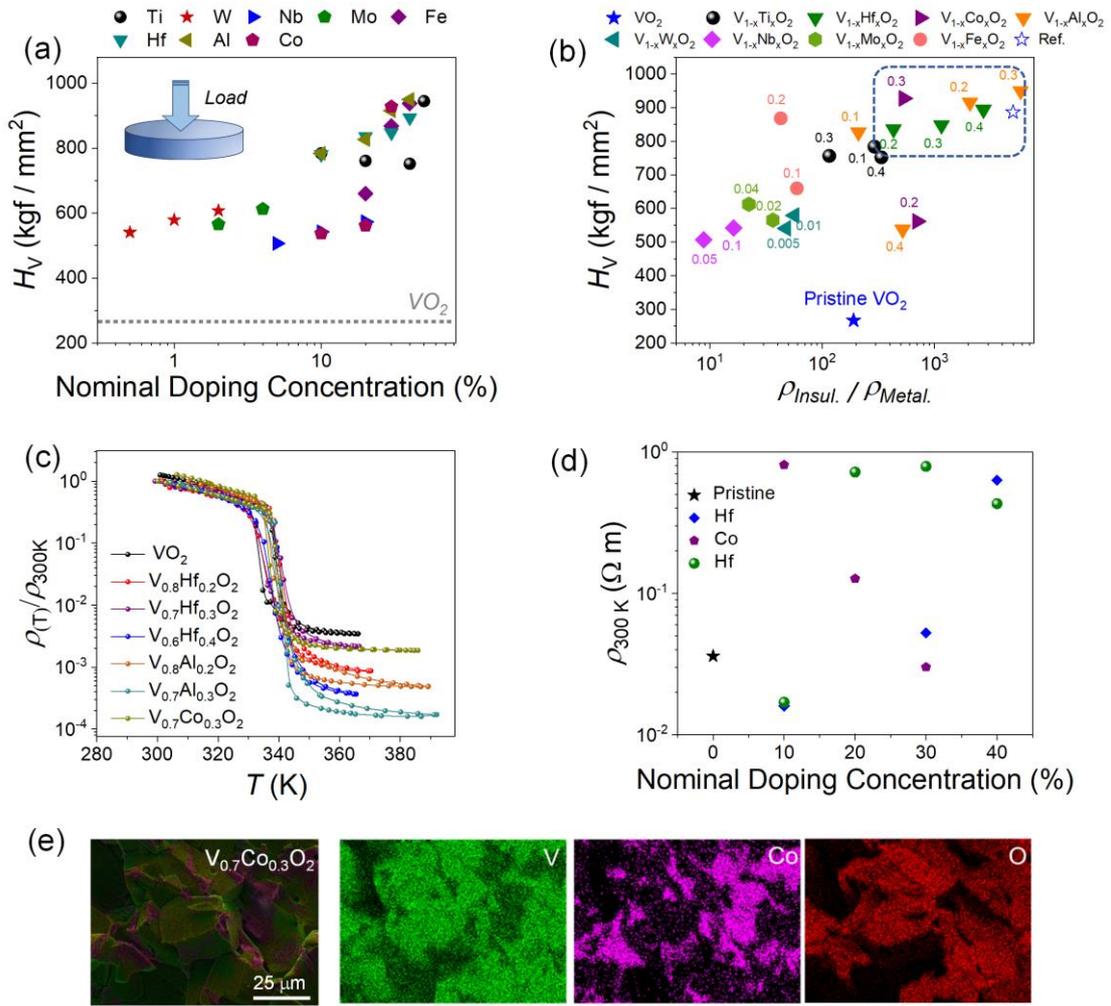

**Fig. 4.** (a) The Vickers hardness ($H_V$) for the transitional metal (*TM*) substituted $VO_2$ (*TM* = Ti, Fe, W, Nb, Mo, Hf, Al and Co) made herein, plotted as a function of the nominal doping concentration. (b) Plotting $H_V$ as a function of the variation in resistivity across the metal to insulator transition temperature (e.g., $\rho_{Insul.} / \rho_{Metal.}$) for as-made $V_{1-x}TM_xO_2$ (the number of x is marked near the data points) pellet, as compared with the ref. 45. (c) The normalized temperature dependence of the resistivity ($\rho$-$T$) of the co-sintered $VO_2$ with transitional metal oxides with high melting points. (d) The basic resistivity at room-temperature of as-made $V_{1-x}TM_xO_2$ plotted as a function of the nominal doping concentration. (e) The energy dispersion spectrum (EDS) of V, Co and O elements for the as-made $V_{0.7}Co_{0.3}O_2$ pellet.